\documentclass[aps,prb,groupedaddress,twocolumn]{revtex4}
\usepackage{epsfig}
\usepackage[dvipsnames,usenames]{color}
\usepackage{color}
\usepackage{amsmath}
\usepackage{comment}
\usepackage{array}
\usepackage{multirow}
\usepackage{tabularx}
\usepackage{float}
\usepackage[colorlinks=true,bookmarks=true, pdfborder={0 0 0},linkcolor=red,urlcolor=red,     breaklinks=true,bookmarksnumbered=true]{hyperref}

\emergencystretch=\maxdimen
\begin{document}
\title{Inversion symmetry breaking in bilayer multi-orbital Hubbard model with impurity approximation}
\author{Chenye Qin and Mi Jiang}
\affiliation{Institute of Theoretical and Applied Physics, Jiangsu Key Laboratory of Thin Films, School of Physical Science and Technology, Soochow University, Suzhou 215006, China}

\begin{abstract}
We investigated the Cu/Ni-O$_2$ bilayer multi-orbital Hubbard model with Cu/Ni impurity approximation embedded in the O lattice by incorporating the 3d$^{8}$ multiplet structure coupled to a full O-2p band. For this simplified model of describing the local electronic structure of hole doped homo-bilayer Cu(Ni)O$_2$-Cu(Ni)O$_2$ and hetero-bilayer CuO$_2$-NiO$_2$, we demonstrate that the homo-bilayer system hosts rich phase diagram with inversion symmetry breaking between two layers at strong enough interlayer hybridization, especially the coexistence of two-hole singlet and triplet states in opposite layers. Besides, we show that the interlayer hybridization in hetero-bilayer systems destabilizes the Zhang-Rice singlets or 3d$^{8}$ triplets within each layer via charge transfer between layers. The relevance of these electronic redistribution to the bilayer or interface systems of cuprates and/or nickelates are discussed.
\end{abstract}

\maketitle

\section{Introduction}
Bilayer single-orbital Hubbard model has been widely adopted to describe the strongly correlated electronic systems, especially on the magnetic and pairing properties~\cite{bilayer1,bilayer2,bilayer3,bilayer4,bilayer5,bilayer6,bilayer7,bilayer8}, as well as exciton condensation~\cite{bilayer9,bilayer10}, and enhancement of superconductivity~\cite{incipient2,incipient3,incipient4,Werner,Maier2022}. Its cousins, e.g. the bilayer Heisenberg~\cite{bilayerHei1,bilayerHei2} and t-J~\cite{bilayertJ1,bilayertJ2} models also serve as the playground for exploring interplay between various quantum phases within each layer, for instance, the antiferromagnetic ordering spin fluctuation mediated pairing, and interlayer physics induced by the hybridization and/or interaction. These work have established the bilayer Hubbard model as the versatile platform for exploring the strongly correlated physics. 

Bilayer models are not simply toy model Hamiltonians given that some cuprate superconductors possess the unit cells consisting of CuO$_2$ bilayer structure~\cite{Keimer2015}. 
One significant ingredient of any bilayer models lies in the interlayer hybridization and/or interaction, which can be simply treated as mimicking the experimental pressure. Interestingly, recent high pressure experiments on cuprates~\cite{Sunliling2022} discovered the quantum phase transition from a superconducting state to an insulating-like state as a function of pressure in Bi$_2$Sr$_2$CaCu$_2$O$_{8+\delta}$ (Bi2212) for a wide range of doping levels regardless of the number of CuO$_2$ planes in a unit cell. 

For the newly discovered Ni-based SC~\cite{2019Nature,Aritareview,Botana_review,Held2022,Hanghuireview}, the latest experimental manipulation of the  NiO$_2$ plane separation through the ionic size fluctuations in the rare-earth spacer layer with different hole dopants revealed that the importance of interlayer physics and the dimensionality of SC in the layered nickelates could be essential to build reasonable theoretical models for accounting for various puzzling phenomena~\cite{Ariando2023}. Therefore, with the advent of newly discovered Ni-based superconductors, it becomes physically plausible to explore the bilayer models potentially relevant to infinite-layer nickelates and even (at least theoretically) the CuO$_2$-NiO$_2$ hetero-bilayer. 

The simplest bilayer single-orbital Hubbard model only depicts the physics originating from $3d_{x^2-y^2}$ orbital, which is widely believed to dominate the cuprate physics~\cite{Anderson87,Anderson872,ZhangRice,Hubreview1,Hubreview2}. The argument of its universality of describing the whole family of cuprate superconductors has long been debated~\cite{nonZhangRice2017,T-CuO,BSCCO_ZRS,Lau,Hadi1, Hadi2}, namely whether other $3d$ orbitals and/or O-2p orbitals, especially those non-planar orbitals~\cite{Feiner1992,Feiner1992a,Feiner1992b,Feiner1992c,Feiner1992d,Feiner1992e,Feiner1992f,
Feiner1992g,Tjeng2003,HideoAoki2012,arpes2018,Jin2018}, which is partly due to the charge transfer insulating nature of the cuprate parent compounds~\cite{GSA1985,Emery}, should also be involved into the proper model of unconventional superconductors.
The most complete study is certainly to include the full multiplet structure
of the Cu, for instance, all singlet and triplet irreducible
representations in the $D_{4h}$ point group spanned by two d holes
($d^8$-type configurations) and their corresponding Coulomb and
exchange interactions\cite{Zaanen1987,Eskes88,Eskes90}. There are a variety of experimental evidence on the multiplet effects, supporting the triplet states arising from the non-planar orbitals due to Hund's rule, e.g. Auger spectroscopy~\cite{Auger1980} and X-ray absorption (XAS)~\cite{Feiner1992} experiments.

\begin{figure} [h!]
\psfig{figure={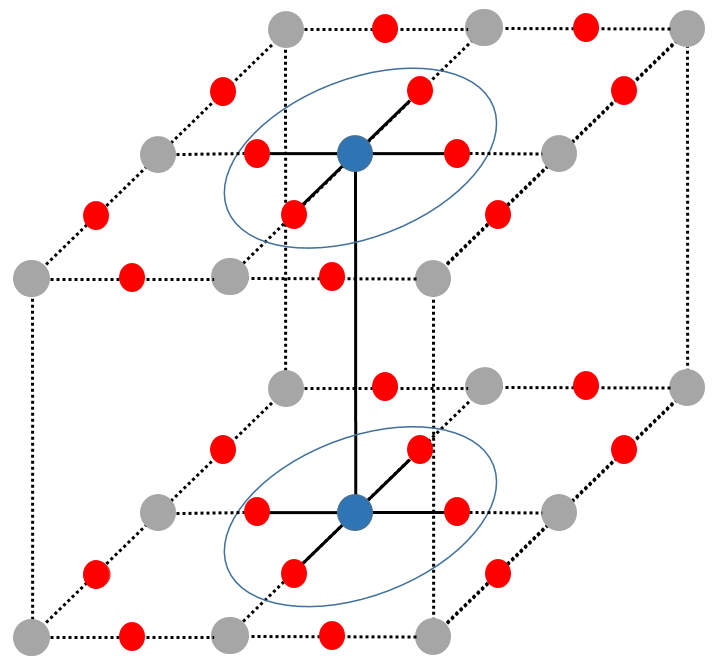},height=5.0cm,width=.4\textwidth, clip} 
\caption{Schematic illustration of the bilayer geometry. All blue and gray dots denote the sites with $d$ orbitals although our study only include the impurity blue sites. Red dots denotes O sublattice. The black solid line denotes the interlayer hybridization and all the dashed lines are neglected hybridizations. The intralayer O-O hybridizations are not shown. The elliptic emphasizes that only the hole states near the two impurities within this region are our focus.}
\label{lat}
\end{figure}

All the above consideration including cuprate and nickelate superconductors motivate us to investigate the bilayer multi-orbital model consisting of the full Cu/Ni-$3d$ multiplet structure and explore its low-energy
properties. There is no doubt that the many-body treatment of such a model in the thermodynamic limit is extremely challenging and even impossible nowadays~\cite{Hubreview1}. 
Hence, in order to obtain numerically exact results, we will follow our previous studies~\cite{Mi2020,Mi2022} to adopt the single Cu/Ni impurity approximation and particularly focus on the effects of the interlayer hybridization on the local electronic structure, namely the original hole-doped states, e.g. Zhang-Rice singlet (ZRS) within each layer.


\section{Model and Method}\label{model}

Our bilayer model is a direct extension of the impurity model adopted previously to study the hole-doped spin state and symmetries relevant for infinite-layer cuprates and nickelates~\cite{Mi2020,Mi2020a,Mi2022}. As shown in Fig.~\ref{lat}, the two impurities locate at the center of each layer with interlayer hybridization. In this work, to avoid the states that the holes preferring to occupy the remote O ions, we restrict on the situation that the interlayer hybridization only exists between the d orbitals of the two impurities. The Hamiltonian reads as follows
\begin{align} \label{H}
H &= E_s + K_{pd} + K_{pp} + V_{dd} + V_{pp} +K_{z} \nonumber \\
E_s &= \sum_{zm\sigma} \epsilon_d(m) d^\dagger_{zm\sigma}d^{\phantom\dagger}_{zm\sigma} 
      + \sum_{zjn\sigma} \epsilon_p(z) p^\dagger_{zjn\sigma}p^{\phantom\dagger}_{zjn\sigma} \nonumber \\ 
K_{pd} &= \sum_{\langle .j\rangle zmn \sigma} 
(T^{pd}_{mn} d^\dagger_{zm \sigma}p^{\phantom\dagger}_{zjn \sigma}+h.c.) \nonumber \\ 
K_{pp} &= \sum_{\langle zjj'\rangle nn' \sigma} 
(T^{pp}_{nn'} p^\dagger_{zjn \sigma}p^{\phantom\dagger}_{zj'n' \sigma}+h.c.) \nonumber \\ 
V_{dd} &= \sum_{\bar{m}_1\bar{m}_2\bar{m}_3\bar{m}_4} U(\bar{m}_1\bar{m}_2\bar{m}_3\bar{m}_4) d^\dagger_{\bar{m}_1}d^{\phantom\dagger}_{\bar{m}_2}d^\dagger_{\bar{m}_3}d^{\phantom\dagger}_{\bar{m}_4} \nonumber \\
K_{z} &= \sum_{m \sigma} 
(T^{z}_{m} d^\dagger_{1m \sigma}d^{\phantom\dagger}_{2m \sigma}+h.c.) 
\end{align}
with the shorthand notation $\bar{m}_x \equiv zm_x \sigma_x$, with $x=1,\dots,4$ denoting spin-orbitals.
Note that this model is reminiscent of our previous studied Hamiltonian. More details on the single layer situation and the formalism on the calculation of the ground state (GS) and the spectra for different irreducible representations of $D_{4h}$ point group are referred to Ref.~\cite{Mi2020,Mi2020a,Mi2022}. 

Here we only remark on the layer $z=1,2$ dependent $\epsilon_p(z)$ for O orbitals and orbital-selective interlayer hybridization, which is our focus in this work.
The two layers have the same (different) $\epsilon_p$ for homo-bilayer (hetero-bilayer). The latter situation can be seen as artificially breaking the inversion symmetry while we show that the former case can induce   inversion symmetry breaking at large enough interlayer hybridization. Note that our model Eq.~\ref{H} assumes the same $\epsilon_d=0$ and interaction $U$ matrix in two layers for simplicity. This is partly motivated by the assumption that the Coulomb and exchange interaction for the infinite-layer cuprate and nickelate superconductors do not differ significantly~\cite{Mi2020,Mi2022,Aritareview,Botana_review,Held2022,Hanghuireview}.


The interlayer hybridization is described by $K_{z}$, where $T^{z}_m$ denotes the d-orbital dependent hybridization, which is usually adopted to mimic the experimental pressure effects. Note that the hybridization only occurs between impurities of two layers and $m\in \{ a_1 \equiv d_{z^2}, b_1 \equiv d_{x^2-y^2}, b_2 \equiv d_{xy}, \}$ owing to the $d$-orbital phases, which also results in relatively larger hybridization between $d_{z^2}$ orbitals compared with $d_{x^2-y^2},d_{xy}$. To capture this difference, we assign a tunable coefficient $\sim 1.2$ to the $d_{z^2}$-$d_{z^2}$ hybridization, which is an approximate value (our numerical calculation indicates that its value in the range of 1.1 to 1.3 induces negligible difference on the presented results).
It is worth mentioning that additionally including the hybridization between O-$2p$ orbitals leads to the states with holes locating at O ions far away from the impurities, which is not our most interested bound states. Therefore, to avoid this wide spectral continuum, we only keep the $d-d$ hybridization. 

One additional complexity arises from the recent experimental demonstration on the significance of the H doping~\cite{H nature}, whose probable occupancy at the apical O position (the same as interstitial s orbital) would block the intercell hybridization between the interstitial s orbital and $d_{x^2-y^2}, p$ orbitals and also influence the onsite hybridization between two NiO$_2$ layers. To account for H doping's impact, we also investigate a simplified model with only $d_{z^2}$-$d_{z^2}$ hybridization between layers since the doped H would strongly hybridize with the nearest $d_{z^2}$ orbital so that there will be an effective interlayer $d_{z^2}$-$d_{z^2}$ hybridization albeit with much weaker magnitude than the situation without H blocking. In other words, in this model, we do not explicitly consider the states associated with doped H, which would further expand the Hilbert space so that complicate the simulations, but rather include it in an implicit manner.

While the single layer problem only requires the treatment of a two-hole problem taking into account the Coulomb and exchange interactions $U(\bar{m}_1\bar{m}_2\bar{m}_3\bar{m}_4)$ for all singlet and/or triplet
irreducible representations of the $D_{4h}$ point group spanned by two $d$ holes~\cite{Mi2020}, the bilayer system consists of four-hole states, which exponentially enlarge the variational Hilbert space. This strongly limits our simulation to relatively small O lattices. Most of our simulations are performed on $4\times 4$ or $6\times 6$ O lattices. Nonetheless, fortunately, the impurity approximation implies that our simulations do not sensitively depend upon the lattice size so that even our relatively small lattice can capture the essential physics discussed here.


Regarding the parameter choice, throughout the paper, the conventional values $A=6.0$ eV, $B=0.15$ eV, $C=0.58$ eV are adopted, where $A$ normally characterizes the strength of the magnitude of electron-electron interaction and here we do not vary it unlike our previous studies~\cite{Mi2020,Mi2020a,Mi2022} following the assumption that the infinite-layer cuprates and nickelates have similar interaction strength. In addition, throughout this work, we use the conventional hybridization $t_{pd} = 1.3$ eV and $t_{pp} = 0.55$ eV. 

Before proceeding, we briefly discuss some possible hole states within our system. In our simulation, we simplified the Hilbert space by restricting the total magnetization of four spins to be $S^z_{tot}=0$, namely there are always two holes with spin up and two others with spin down. Note that although the total spin of four holes can be $S_{tot}=0,1,2$,  in this work we mainly quantify the intralayer two-hole total spin by singlet or triplet but neglect the information of interlayer spin states or bonding/antibonding states. 

Except for the interlayer hybridization, another major control parameter is the charge transfer energies $\epsilon_p$ of two layers. On the one hand, if each layer has the same $\epsilon_p$, namely the homo-bilayer, then at relatively weak interlayer hybridization, four holes would evenly distribute with  exactly the same two-hole states in each layer. In other words, the homogeneity between two independent layers would persist. One goal of this work is to provide evidence that this homo-bilayer will spontaneously break the inversion symmetry between layers at sufficiently large hybridization.
On the other hand, for the hetero-bilayer, the aforementioned inversion symmetry between layers is artificially broken, which is realized via distinct $\epsilon_p$ of two layers, so that the holes preferentially locate on the layer with smaller $\epsilon_p$ such that there will be states with one hole on one layer and three holes on the other.

To label the four-hole states, we use the notation such as $d^8$-$d^8$ or detailed $b_1b_1$-$a_1b_1$, $b_1L$-$a_1L$, and $b_1$-$b_1L^2$ etc. to denote the state on each layer separated by hyphen. Note that for hetero-bilayer, we adopt the convention that the state before/after the hyphen corresponds to the layer with larger/smaller $\epsilon_p$.

\begin{figure*} [t]
\psfig{figure=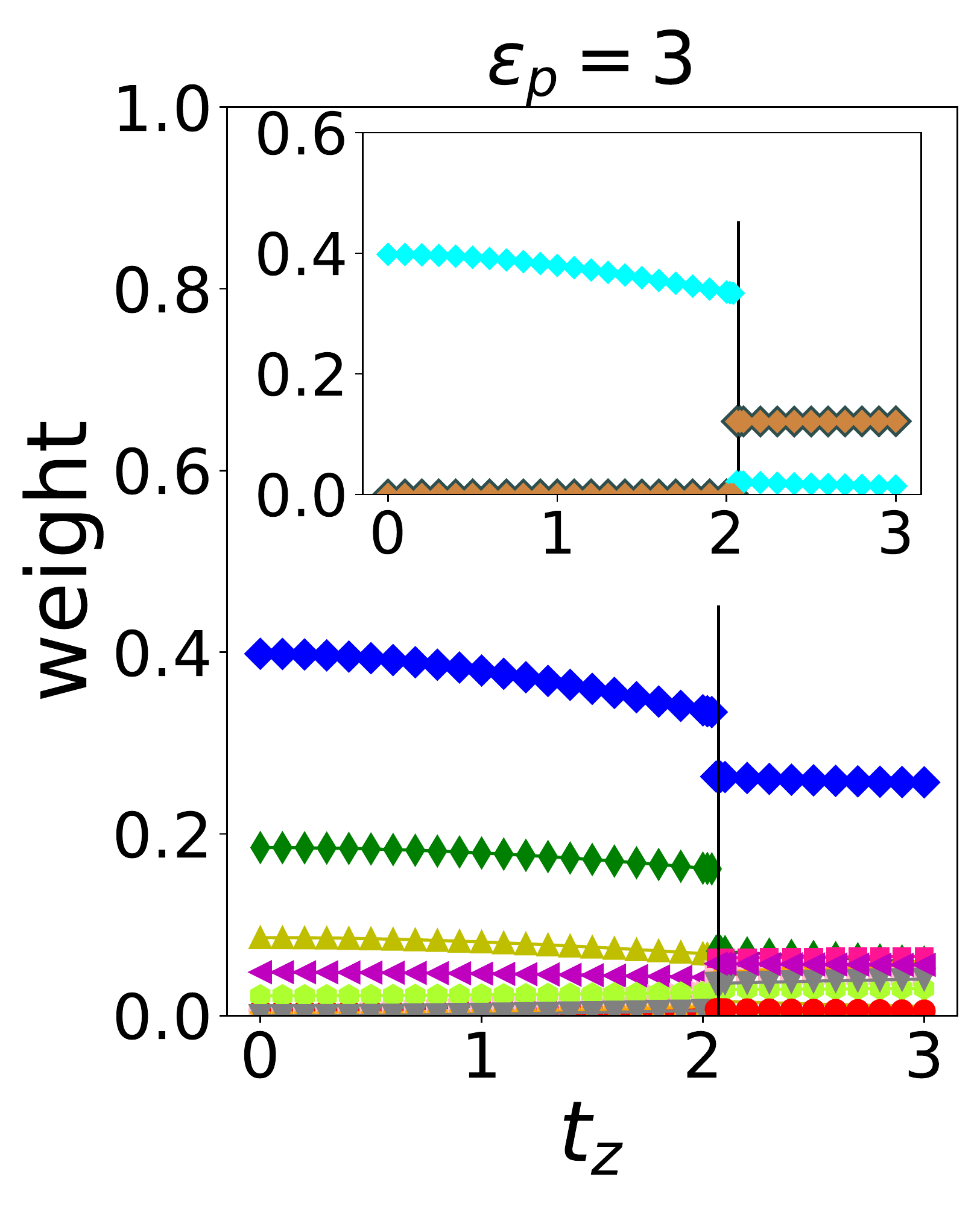},height=7.4cm,width=.34\textwidth, clip} 
\psfig{figure=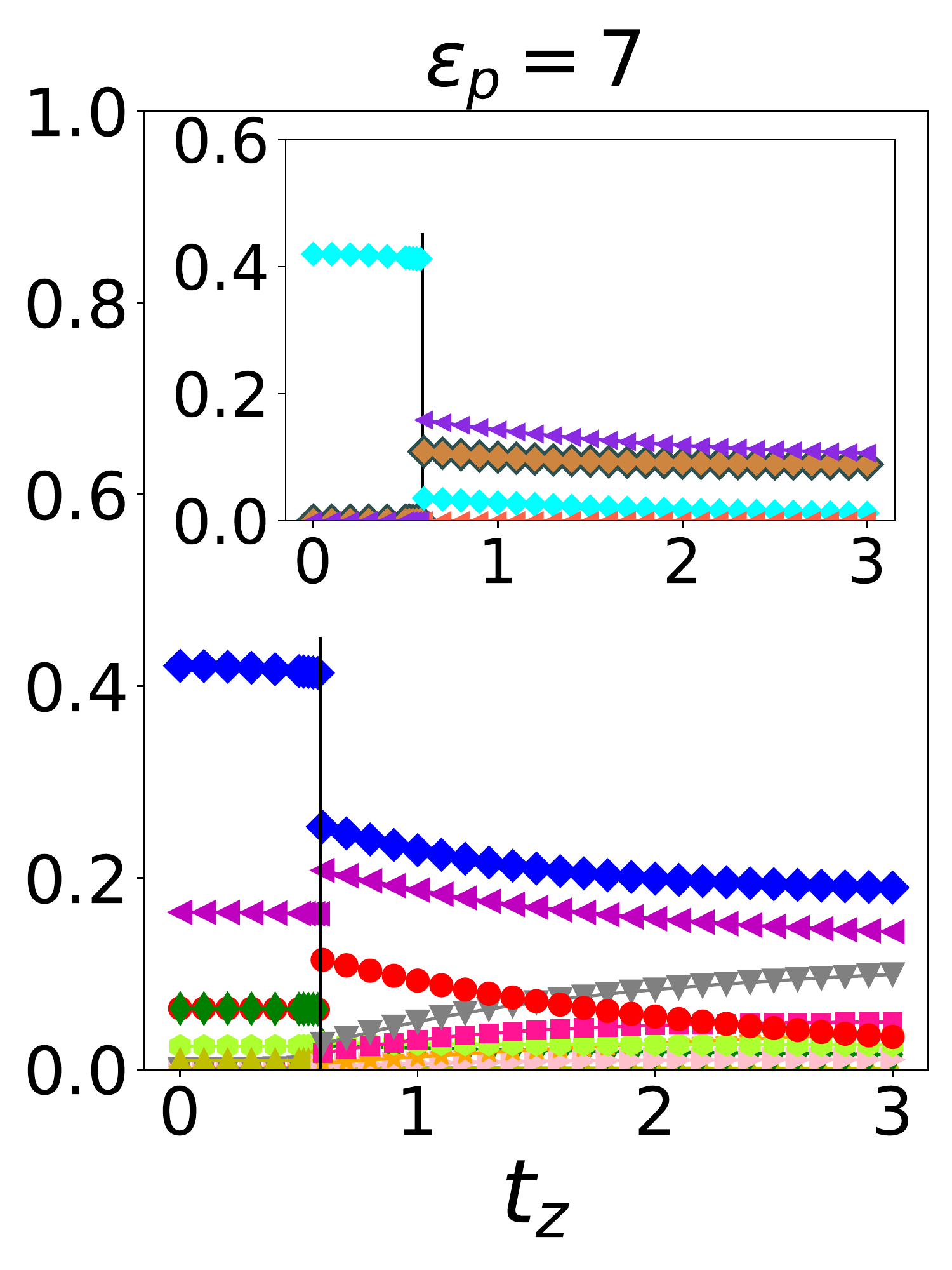},height=7.4cm,width=.31\textwidth, clip} 
\psfig{figure=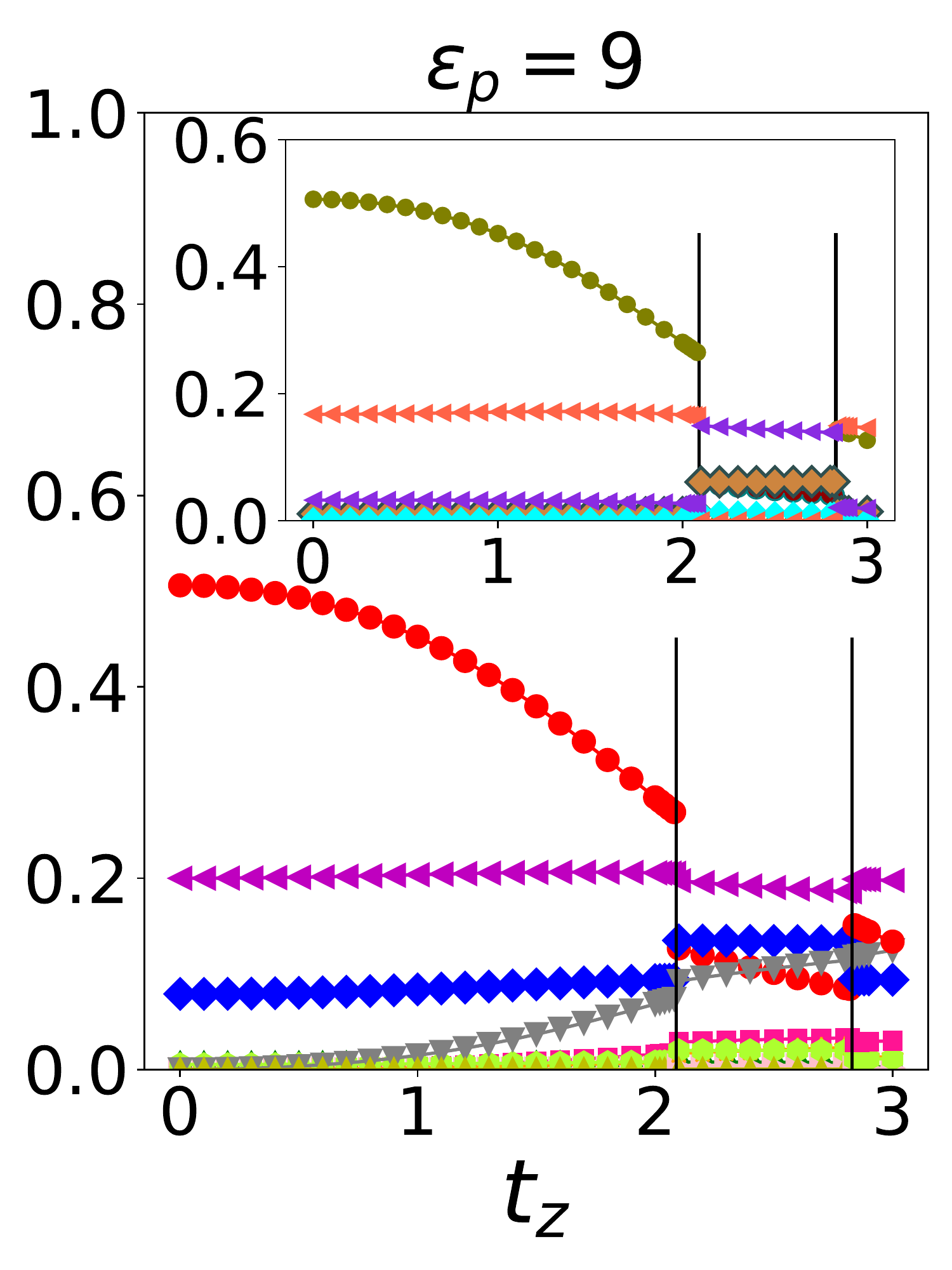},height=7.4cm,width=.31\textwidth, clip}
\psfig{figure={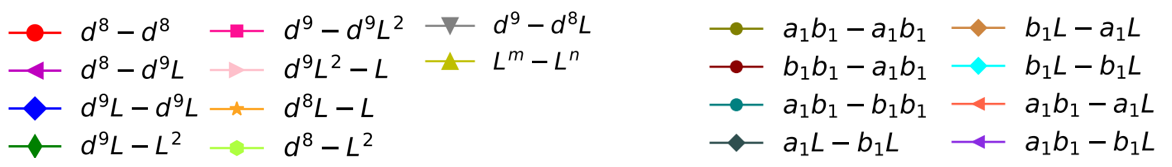},height=2cm,width=.8\textwidth, clip} 
\caption{Evolution of the GS weight distribution with increasing interlayer hybridization $t_z$ between all $d$ orbitals of homo-bilayer model for three characteristic $\epsilon_p=3,7,9$ eV. The common observation is the   inversion symmetry breaking at large $t_z$, where the GS involves strong $d_{z^2}$ components tending to form two-hole triplets.}
\label{fig1}
\end{figure*}

\section{Results}\label{results}
In the monolayer situation~\cite{Mi2020}, $\epsilon_p=3$ eV that is typical for cuprates locates the system as Zhang-Rice singlet (ZRS), which is characterized by the dominant $b_1L$ weight. For $\epsilon_p=7$ eV, which is estimated value corresponding to the infinite-layer nickelates~\cite{Mi2022}, the ground state (GS) is still readily ZRS but with significant $b_1b_1$ weight. Further increasing $\epsilon_p$, such as $\epsilon_p=9$ eV, induces the phase transition of GS from ZRS to $a_1b_1$ triplet state promoted by Hund's rules. With the above brief reiteration of our previous work on the monolayer, next we explore various bilayer systems.

\subsection{Homo-bilayer: all $d$-orbital hybridization}

We first discuss the homo-bilayer with interlayer hybridization between all $d$ orbitals. Fig.~\ref{fig1} illustrates the evolution of the GS weight distribution with increasing $t_z$ for aforementioned three characteristic $\epsilon_p=3,7,9$ eV. All three panels shows that $t_z$ can have minor effects on weights of various states in quite a wide range; while the weights have drastic jump across the GS phase transition. The insets provide more details of the weight distribution of dominant states like $d^8$ and/or $d^9L$. 
Specifically, panel (a) for $\epsilon_p=3$ eV corresponding to bilayer cuprates shows that turning on $t_z$ only gradually decreases the $b_1L$-$b_1L$ weight until the abrupt transition at $t_z \sim 2.1$ (marked by solid black vertical line) beyond which the dominant state becomes $a_1L$-$b_1L$. Note that $a_1L$-$b_1L$ is degenerate with its parity $b_1L$-$a_1L$ state, namely their weights in GS are exactly the same. The remarkable observation is that the GS is not degenerate (see also discussion on Fig.~\ref{dege} below) so that the inversion symmetry between homo-bilayer is  spontaneously broken by the interlayer hybridization, which constitutes our main finding here. Another important observation is that $a_1L$ is of triplet nature while $b_1L$ is of singlet as expected. In fact, for all models throughout this work, once the $d^8$ or $d^9L$ state involves $a_1$ orbital, the two-hole state is of triplet nature.

Switching to panel (b) for $\epsilon_p=7$ eV roughly corresponding to bilayer nickelates~\cite{Mi2020}, one distinct difference from panel (a) is the phase boundary shifts towards weaker hybridization $t_z \sim 0.6$. In other words, at larger $\epsilon_p$, the GS is easier to be affected by $t_z$, which is connected to the monolayer situation where $\epsilon_p=7$ is close to the phase transition from two-hole singlet to triplet GS so that relatively weak $t_z$ might promote this transition. Another apparent difference is that the transited state at large $t_z$ has substantial weights from $d^8$-$d^9L$, precisely $a_1b_1$-$b_1L$ as shown in the inset. Because of its parity states with the same weight, the dominant GS state would be $a_1b_1$-$b_1L$ although the main panel (b) shows that its sole weight is smaller than $d^9L$-$d^9L$. Obviously, here the inversion symmetry is also broken after the phase transition.

Panel (c) corresponds to a limiting large $\epsilon=9$ case, where the monolayer possesses two-hole triplet GS. Turning on interlayer hybridization rapidly suppresses the weight associated with $a_1b_1$-$a_1b_1$ (see inset) and induces double transitions: the first from $a_1b_1$-$a_1b_1$ with inversion symmetry to an intermediate GS of $a_1b_1$-$b_1L$ similar to the case of $\epsilon=7$ in panel (b) but the second transition to dominant $a_1b_1$-$a_1L$ with both triplets. Hence, large enough $t_z$ prefers $a_1$ orbital to form triplet state with $b_1$ or $L$ holes.
Interestingly, all weights before the first and after the second transition seem to smoothly connect to each other while interrupted by the intermediate GS between the two transitions. 

\begin{figure*} [t!]
\psfig{figure={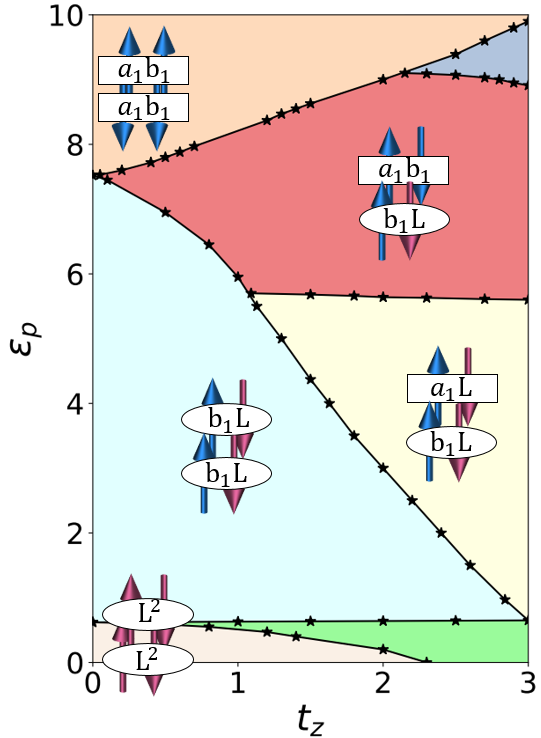},height=9cm,width=.47\textwidth, clip} 
\psfig{figure={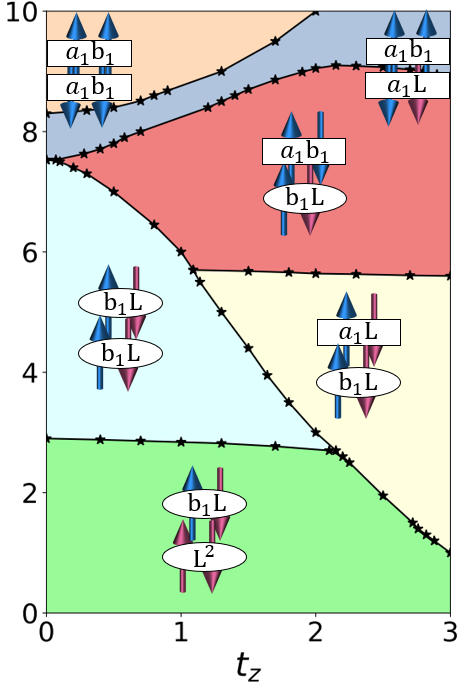},height=9cm,width=.43\textwidth, clip} 
\caption{Phase diagram of homo-bilayer with varying interlayer hybridization $t_z$ between all $d$ orbitals and $\epsilon_p$. The blue and red arrows denote the spin direction of $d$ and ligand hole separately. The elliptic and rectangle labels the two-hole state within one layer as singlet and triplet respectively.}
\label{phase1}
\end{figure*}

To fully capture the physical picture in this homo-bilayer system, Fig.~\ref{phase1} illustrates the phase diagram scanning a wide range of $t_z$ and $\epsilon_p$. The blue and red arrows denote the spin direction of $d$ and ligand hole separately. The elliptic and rectangle labels the two-hole state within one layer as singlet and triplet respectively. The left panel gives the phase diagram without considering the exactly same weight of parity states after breaking the inversion symmetry between layers; while the right panel renormalizes the phase diagram by summing over two degenerate parity states, which does not modify any essential physics.

At $t_z=0$, there are two critical $\epsilon_p$ separating the monolayer two-hole state, specifically from $L^2$ to well-known ZRS $b_1L$ state and finally $a_1b_1$ triplet states. The rich phase diagram displays various transitions between two-hole singlets and triplets. Importantly, the common thread is that $t_z$ tends to involve strong $d_{z^2}$ components that form two-hole $d^8$ or $d^9L$ triplets within each layer. Therefore, both large $t_z$ and $\epsilon_p$ preferentially support the formation of two-hole triplets. 

Generally, relatively weak interlayer hybridization does not strongly modify the GS nature except around critical points such as $\epsilon_p\sim 7.5$; while sufficiently large $t_z$ breaks the inversion symmetry between two layers as discussed in Fig.~\ref{fig1}. As mentioned before, the GS is not degenerate but consists the parity states with the same weight. 
Corresponding to Fig.~\ref{fig1}(c), the phase diagram indicates that the system hosting two successive transitions only occurs for a short range of $\epsilon_p\sim 8-9$, where the intermediate GS of $a_1b_1$(triplet)-$b_1L$(singlet) between two transitions is seemingly unstable and will finally transits to $a_1b_1$(triplet)-$a_1L$(triplet).

Since the interlayer hybridization can generically mimic the effects of high pressure experiments and given that $\epsilon_p\sim 3, 7$ eV are relevant to cuprates and nickelates respectively~\cite{Mi2020a,Aritareview,Botana_review}, Fig.~\ref{phase1} suggests that the pressure experiments for both systems might see the involvement of $a_1 \equiv d_{z^2}$ orbital in the low-energy physics. 
This might be relevant to the recent discovery of the $T_c$ suppression SC-insulator transition~\cite{Sunliling2022}.

Besides, the large $\epsilon_p$ homo-bilayer is prone to be influenced by $t_z$, namely relatively weak pressure on infinite-layer nickelates might induce drastic observable effects. Nonetheless, the appearance of triplet $a_1b_1$ within one layer (red colored phase) is normally detrimental to the superconductivity~\cite{Mi2020a}. This contradicts with the recent report of the pressure induced monotonic enhancement of $T_c$~\cite{pressure}, which probably originates from that the shrunk c-axis lattice constant by pressure will enhance the hybridization between the Ni-3d and rare-earth's 5d orbitals. Because our current investigation only consider NiO$_2$ plane, it possibly further support the idea that those rare-earth's orbitals are crucial for the understanding of the realistic nickelates~\cite{Mi2022,Aritareview,Botana_review,Held2022,Hanghuireview}. In contrary, at smaller $\epsilon_p \sim 3$ eV relevant to cuprates, the transition to the $a_1L$ triplet in one layer requires quite strong $t_z=2$. 
Because of the ongoing debate on the role of $a_1$ orbital~\cite{Mi2020} and also newly discovered cuprate superconductor Ba$_2$CuO$_{4-\delta}$ where the $a_1$ orbital can be essential~\cite{Jin2018}, our phase diagram provides some valuable insights on the role of the interlayer hybridization on promoting the importance of $a_1$ orbital. 

Finally, we remark that the right panel of Fig.~\ref{phase1} provides a supplementary phase diagram accounting for the exactly same weight of parity states after  inversion symmetry breaking. Precisely, if the weights of these states are doubled, then their total weight can exceed those states preserving the inversion symmetry so that the phase diagram will be modified accordingly. We emphasize that this supplementary diagram does not qualitatively change any essential physics.
Clearly, for small $\epsilon_p$ and also around $\epsilon_p \sim 8$, turning on $t_z$ immediately breaks the inversion symmetry, which is hence a generic consequence of the interlayer hybridization in homo-bilayer.


\begin{figure*} [t]
\psfig{figure=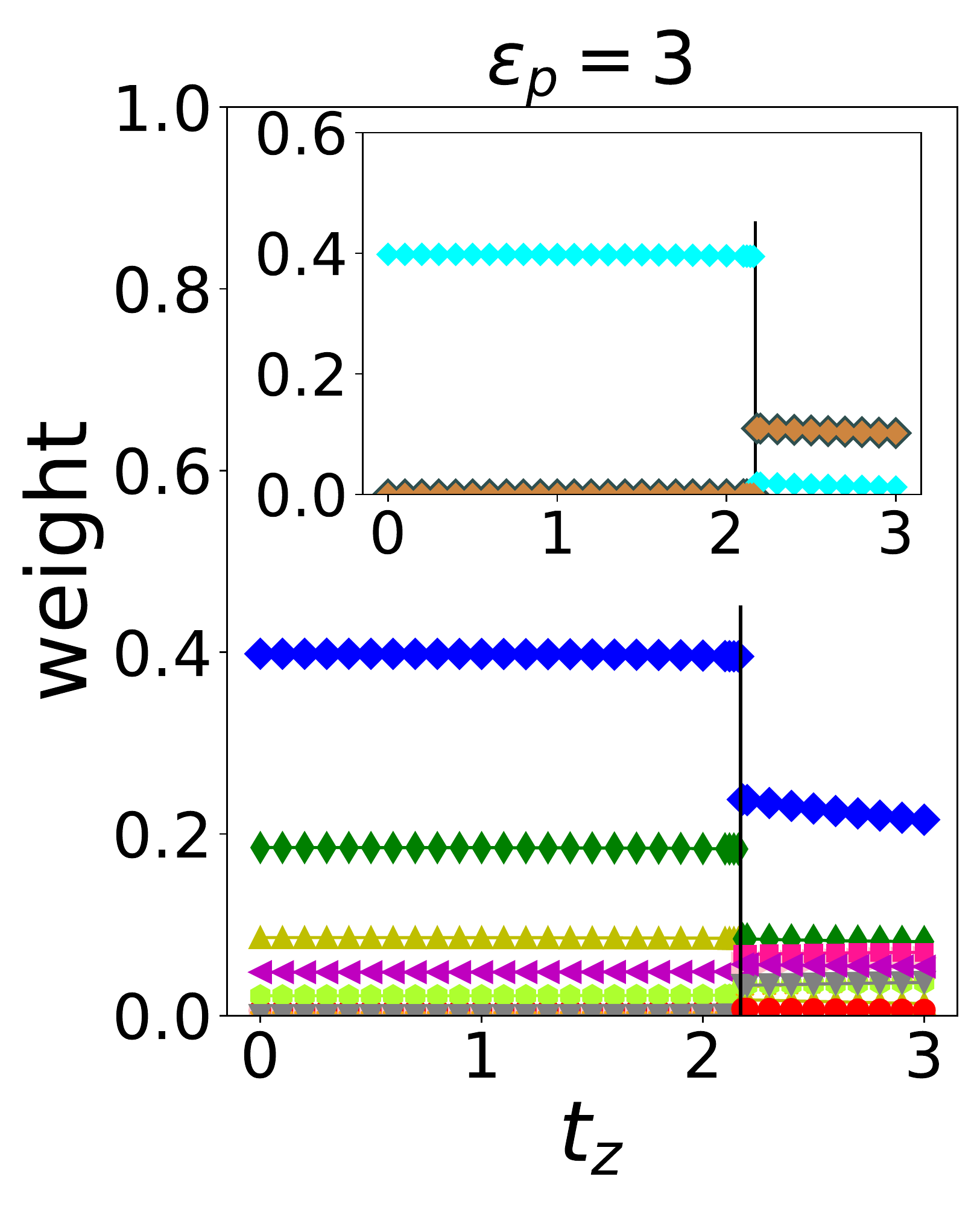},height=7.3cm,width=.34\textwidth, clip} 
\psfig{figure=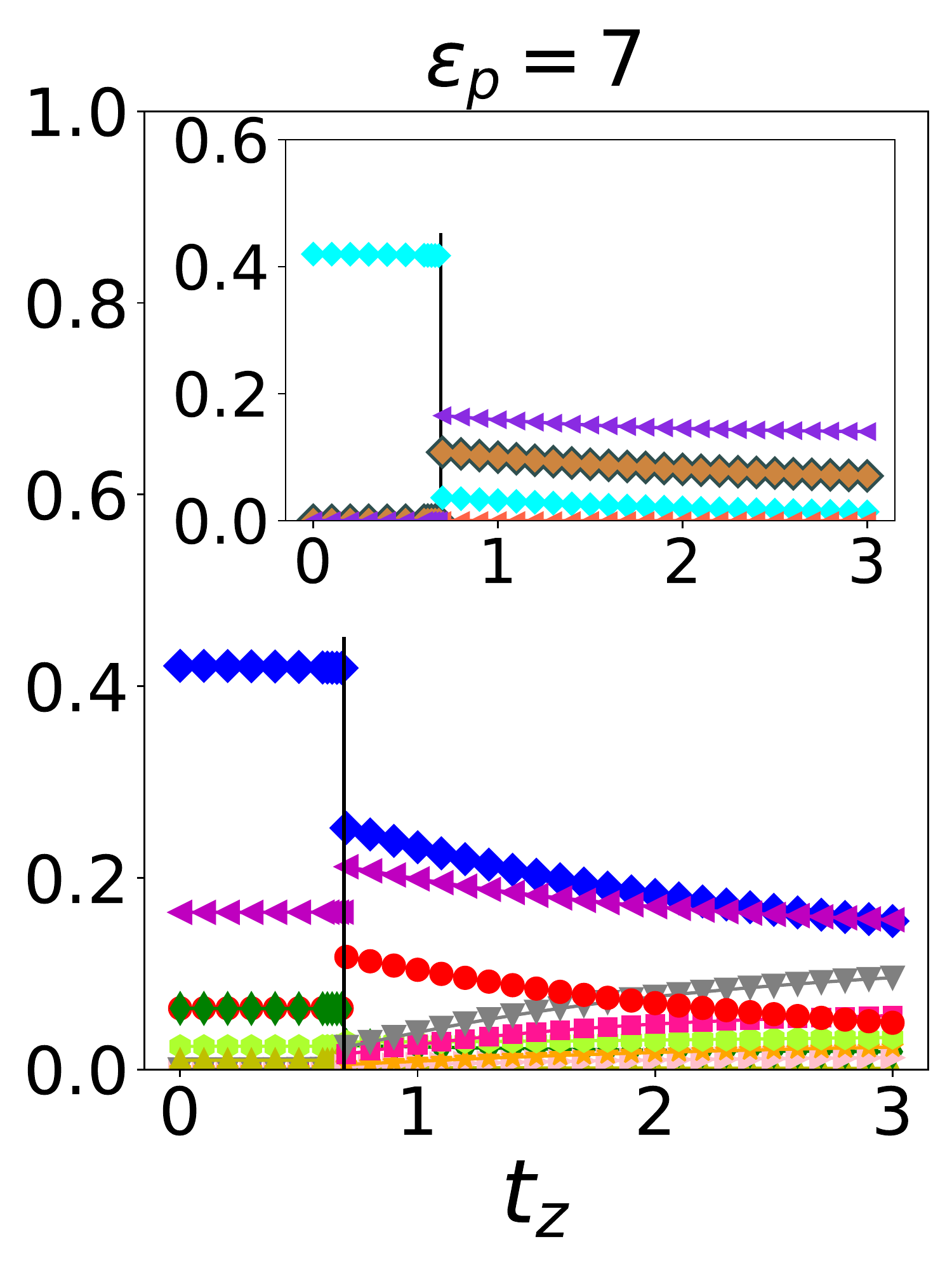},height=7.3cm,width=.31\textwidth, clip}
\psfig{figure=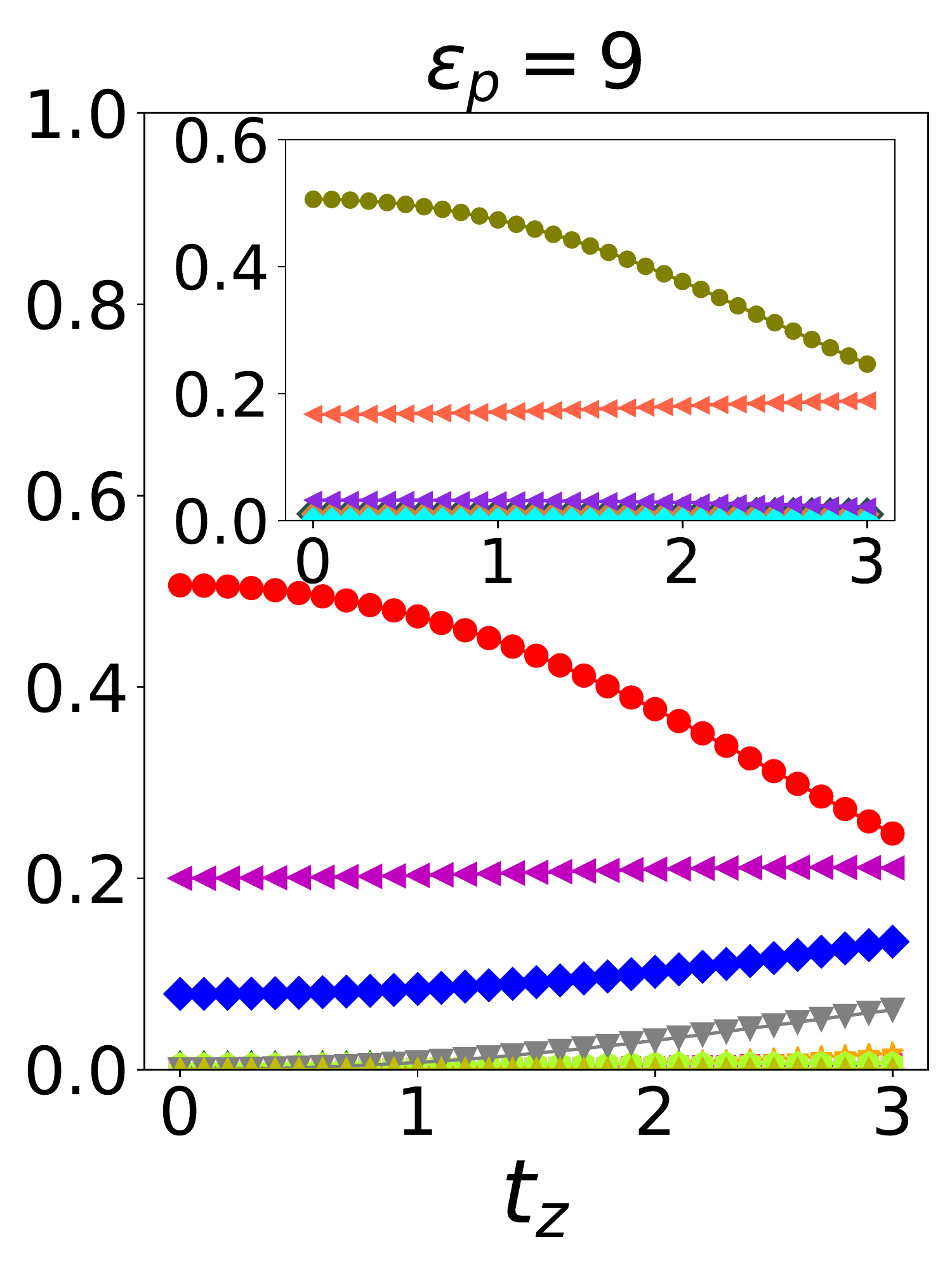},height=7.3cm,width=.31\textwidth, clip} \\
\psfig{figure={./fig2d.png},height=2cm,width=.8\textwidth, clip} 
\caption{Evolution of the GS weight distribution with increasing interlayer hybridization $t_z$ between only $d_{z^2}$ orbitals of homo-bilayer model for three characteristic $\epsilon_p=3,7,9$ eV, similar to Fig.~\ref{fig1}.}
\label{figdz2}
\end{figure*}

\subsection{Homo-bilayer: $d_{z^2}$-orbital hybridization}

\begin{figure*} [t!]	
\psfig{figure={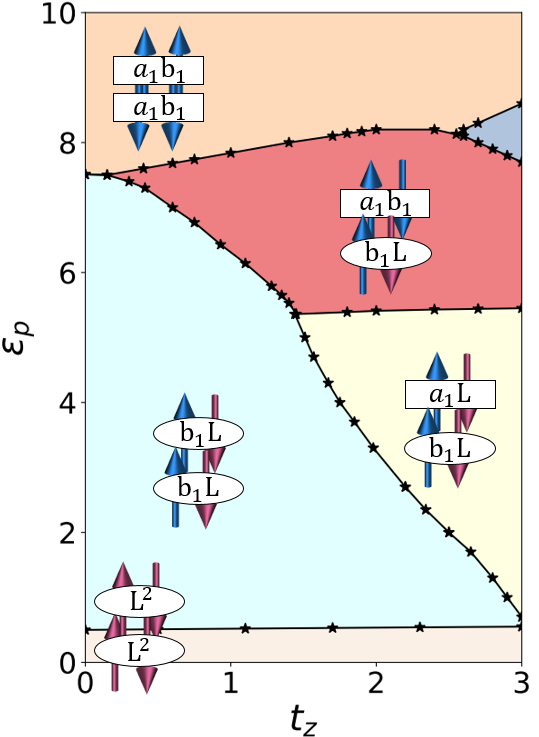},height=9cm,width=.47\textwidth, clip} 
\psfig{figure={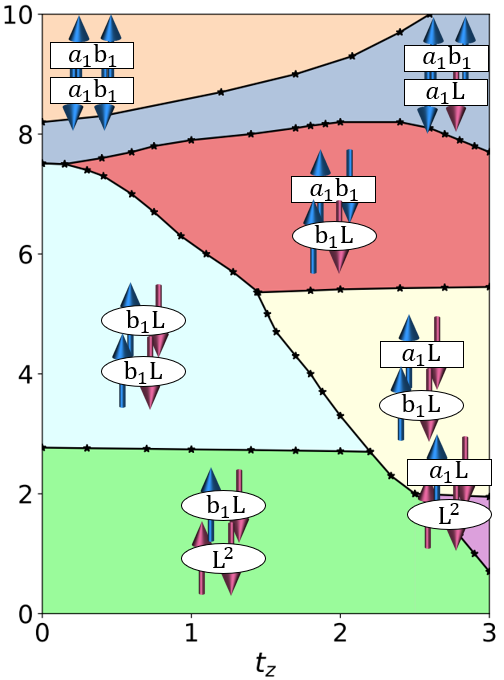},height=9cm,width=.43\textwidth, clip} 
\caption{Phase diagram of homo-bilayer with only $d_{z^2}$-$d_{z^2}$ interlayer hybridization akin to Fig.~\ref{phase1}.}
\label{phase2}
\end{figure*}

We have mentioned the recent experimental demonstration of H doping's impact so that we also investigated a simplified model with only $d_{z^2}$-$d_{z^2}$ hybridization between layers without considering the electronic state of H$^-$ explicitly~\cite{H nature}. In reality, $d_{z^2}$-$d_{z^2}$ is an effective hybridization originating from the strong hybridization between the doped H with the nearest $d_{z^2}$ orbital. In the context of our model, it is natural to expect that larger $t_z$ would be desired to break the inversion symmetry owing to the lack of those additional hybridization from other $d$ orbitals.
As shown in Fig.~\ref{figdz2} corresponding to the three characteristic $\epsilon_p=3,7,9$ eV similar to Fig.~\ref{fig1}, the critical $t_z$ only shifts towards slightly higher values for $\epsilon_p=3,7$ while disappears for $\epsilon_p=9$ in the shown $t_z$ range. In other words, the interlayer hybridization between $d$ orbitals other than $d_{z^2}$ only induce significant effects when the monolayer GS is of triplet nature at large $\epsilon_p$. In fact, the double transition observed in Fig.~\ref{fig1} still exists but at smaller $\epsilon_p$ as supported in the phase diagram Fig.~\ref{phase2}.

Further examination reveals that the dominant feature for $\epsilon_p=3$ (panel a) is the robustness of all the weights against $t_z$, namely almost completely independent on $t_z$. Besides, the absence of double transitions in the case of $\epsilon_p=9$ (panel c) further reveals the immunity of the homo-bilayer system with the presence of only $d_{z^2}$-$d_{z^2}$ hybridization.  
This robustness potentially serves as the evidence that H blocking preserves the quasi-2D nature of the infinite-layer nickelates, which is claimed to be beneficial to the superconductivity~\cite{H nature}. 

Despite of these detailed difference, the two homo-bilayer models with different forms of interlayer hybridization share the common features regardless of phase boundary shift.
As shown in the phase diagram Fig.~\ref{phase2} displaying the interplay between $d_{z^2}$-$d_{z^2}$ hybridization and $\epsilon_p$. The essential features in terms of possible phases and their transitions maintain compared with Fig.~\ref{phase1}, where the  major difference is simply the shift of the phase boundaries.

\begin{figure*} [t!]
\psfig{figure=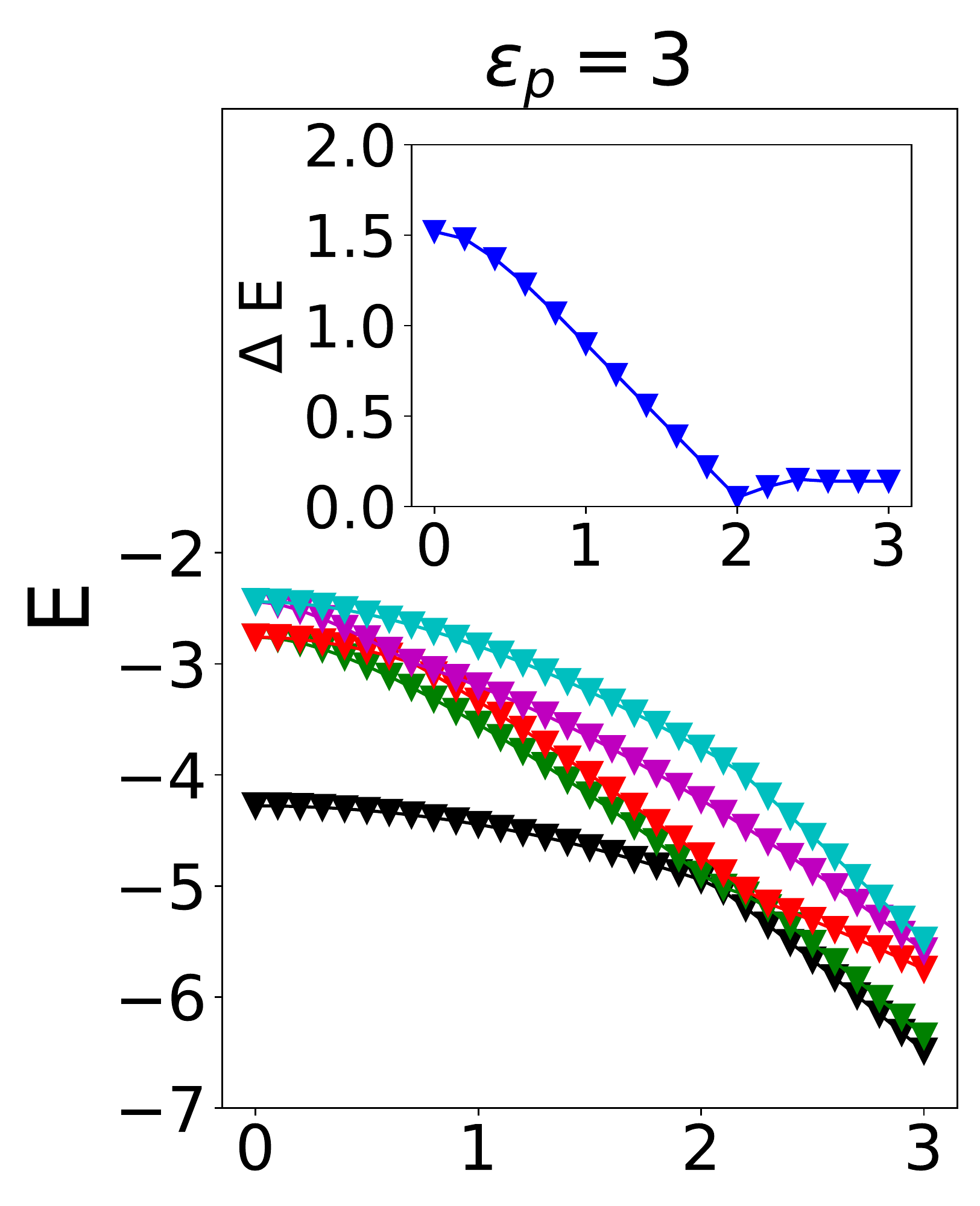},height=7.4cm,width=.34\textwidth, clip} 
\psfig{figure=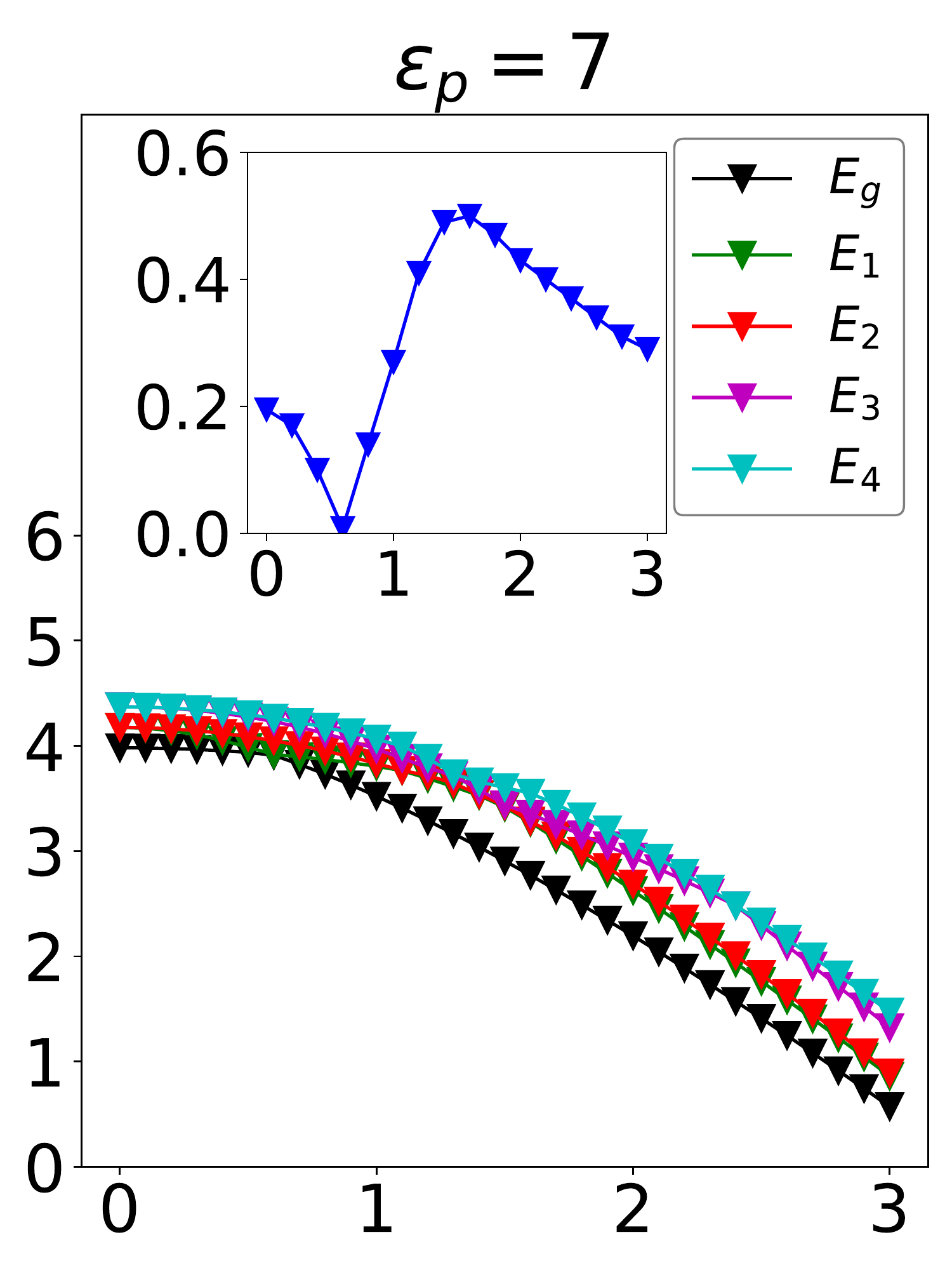},height=7.4cm,width=.31\textwidth, clip} 
\psfig{figure=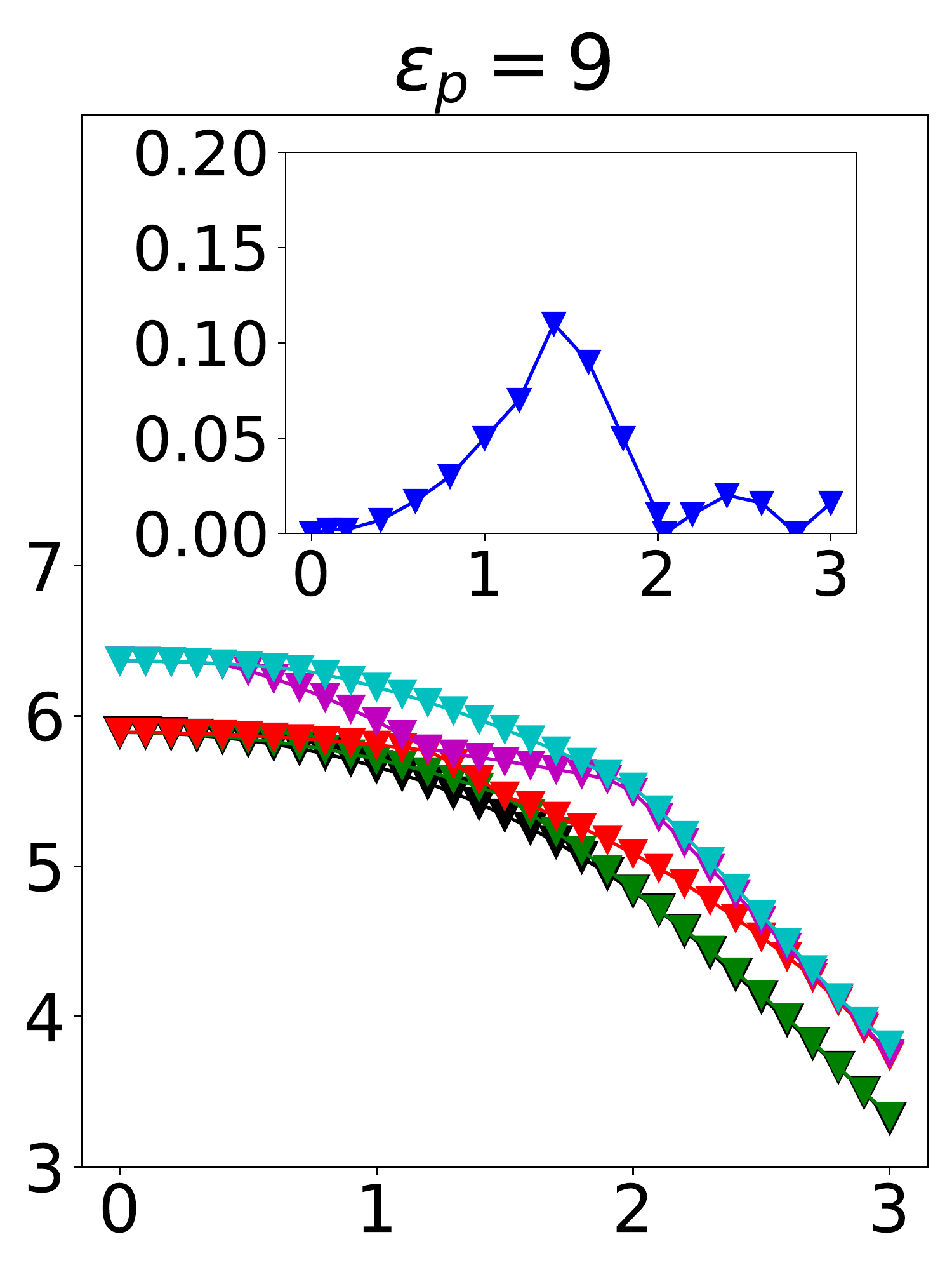},height=7.4cm,width=.31\textwidth, clip} 
\psfig{figure=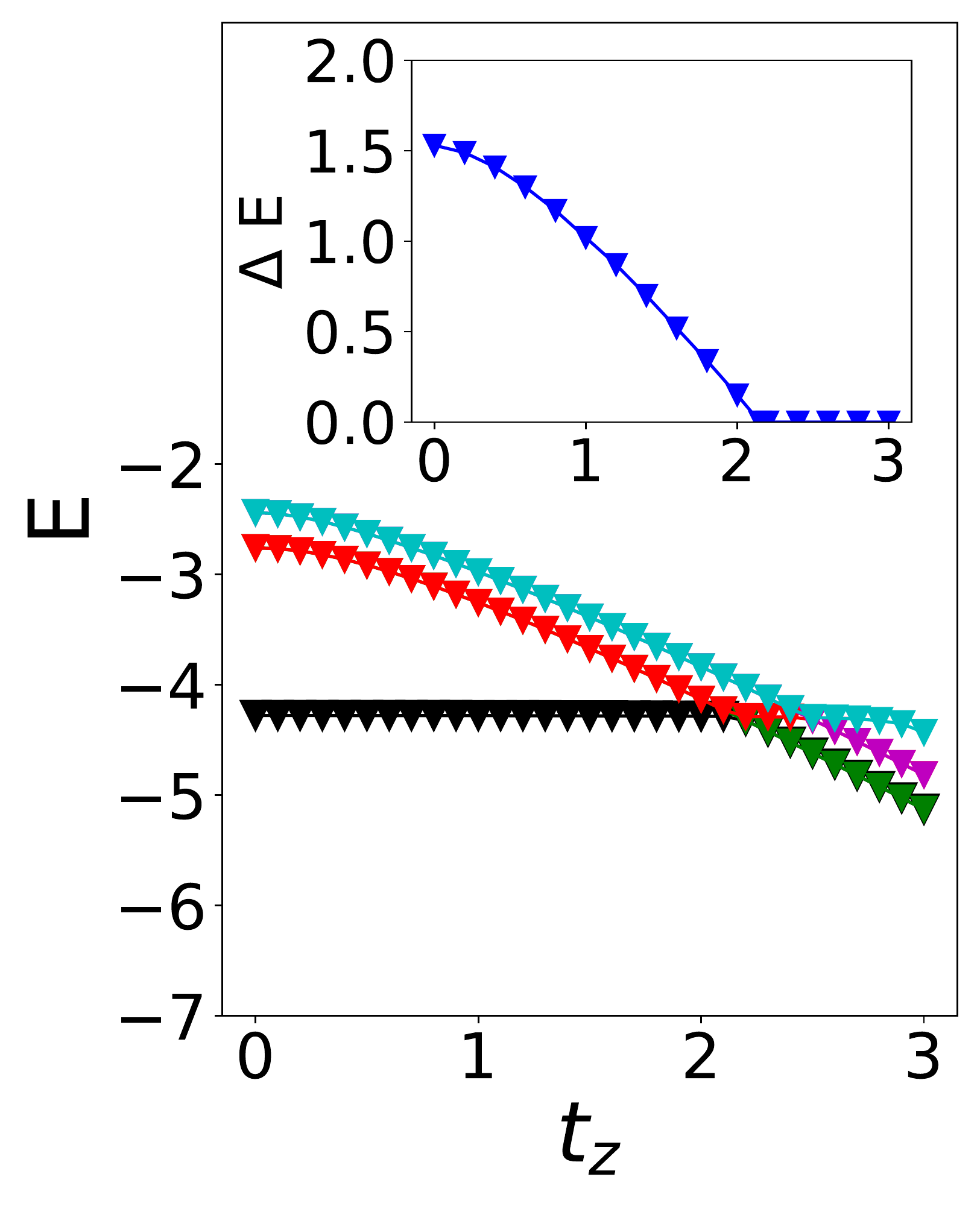},height=7.4cm,width=.34\textwidth, clip} 
\psfig{figure=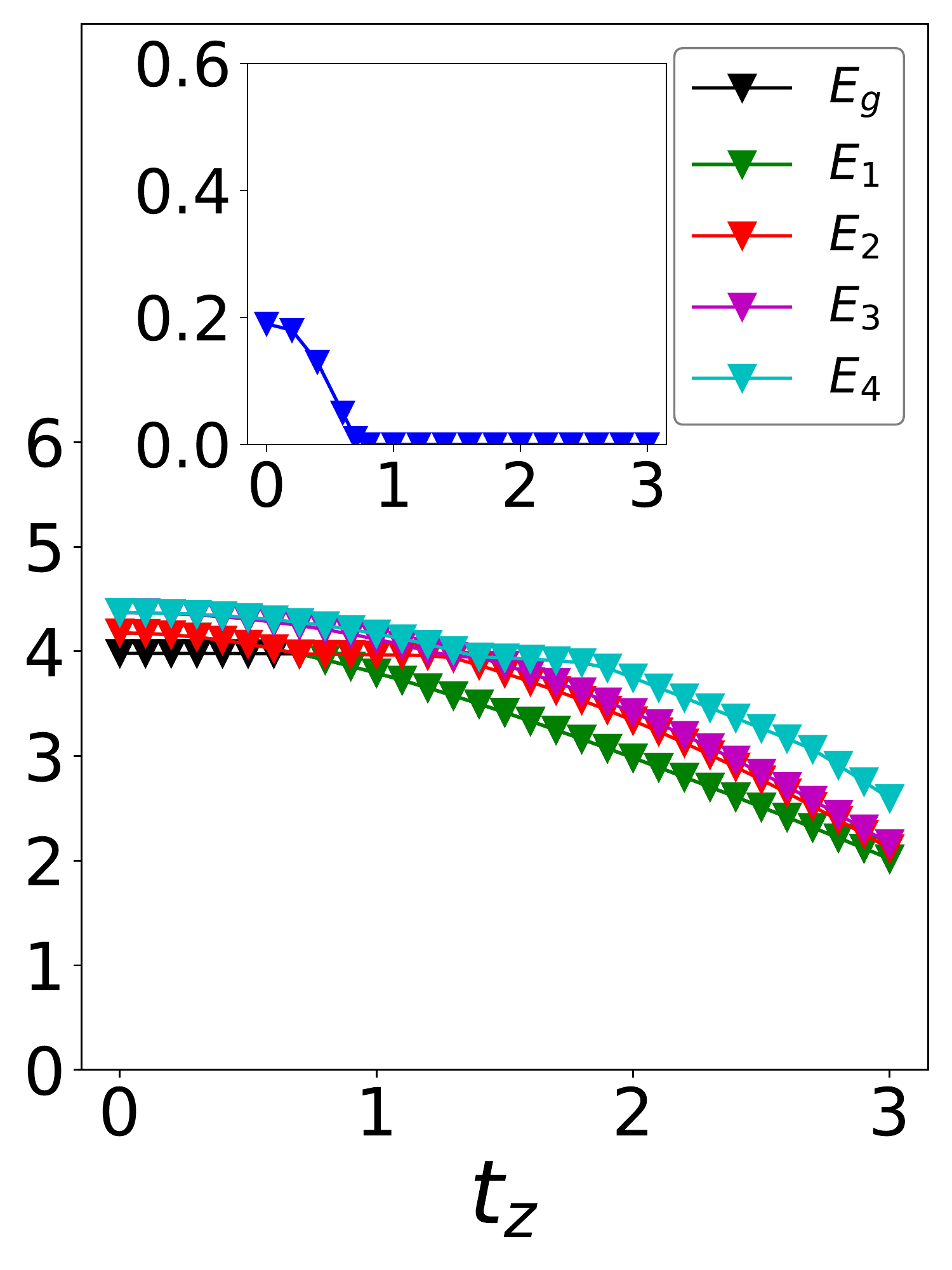},height=7.4cm,width=.31\textwidth, clip} 
\psfig{figure=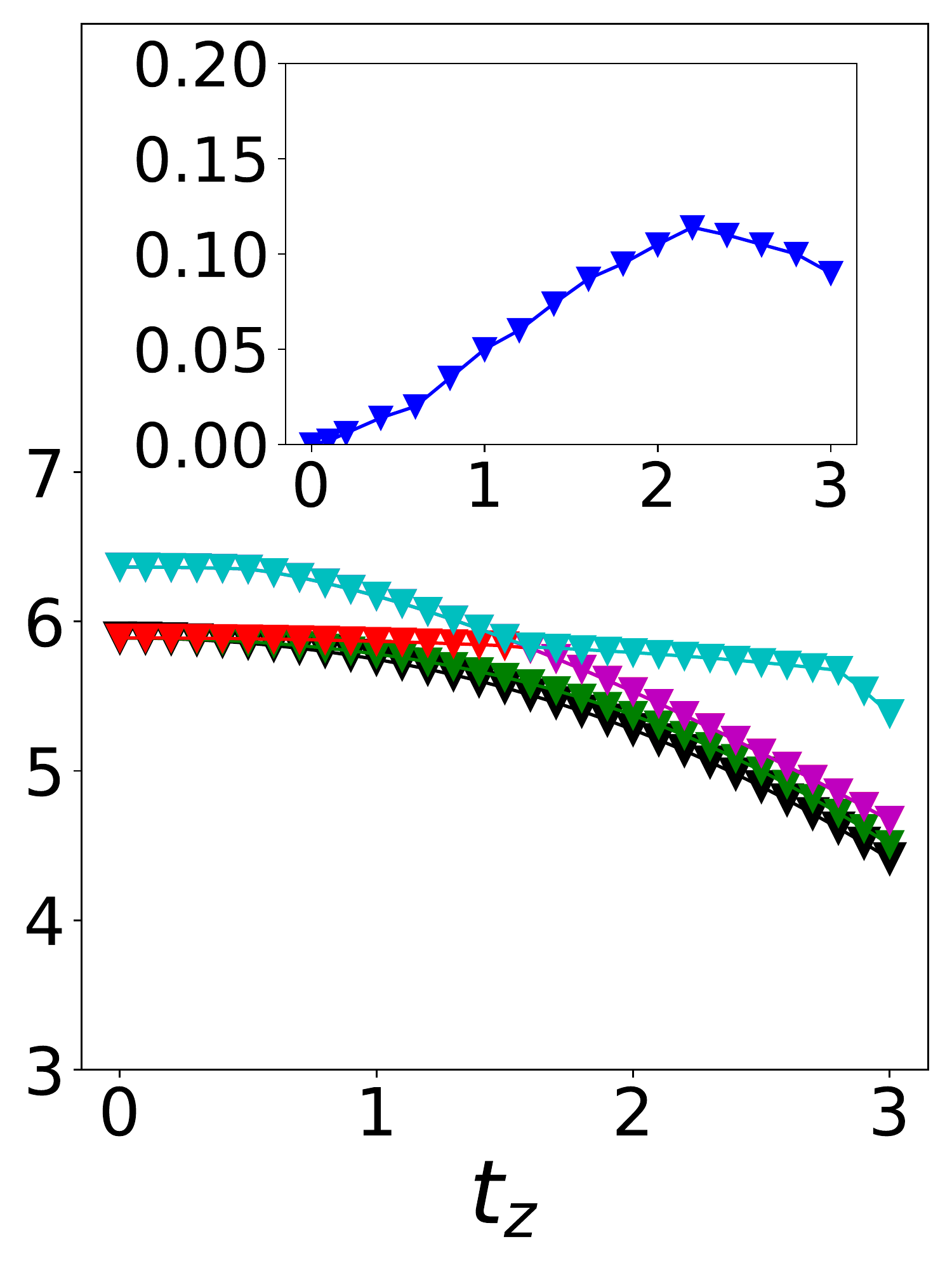},height=7.4cm,width=.31\textwidth, clip} 
\caption{Comparison of the lowest five energy levels $E_i, i=0,1,2,3,4$ with $E_g=E_0$ between two homo-bilayer models with (upper) all $d$ and (lower) only $d_{z^2}$-$d_{z^2}$) interlayer hybridization corresponding to $\epsilon_p=3,7,9$ eV from left to right. The inset shows the ``energy gap'' $\Delta E \equiv E_1-E_g$, which manifests drastic difference between models corresponding to the GS degeneracy.}
\label{dege}
\end{figure*}

Although the GS weights and phase diagrams between two models are seemingly quite similar, we further discovered that the GS degeneracy can host distinct discrepancy between two models. To investigate the GS degeneracy, Fig.~\ref{dege} compares the lowest five energy levels $E_i, i=0,1,2,3,4$ with $E_g=E_0$ between two models corresponding to $\epsilon_p=3,7,9$ eV from left to right. Note that $\epsilon_p$ significantly enhances the energy scales of all the levels.
Again, the model with only $d_{z^2}$-$d_{z^2}$ interlayer hybridization (lower panels) shows generally stronger robustness of $E_i$ against $t_z$, namely its range of variation is smaller. Specifically, as shown in the separate five curves in the upper panels compared with the degenerate three curves in the lower panels, the degeneracy between levels are easier to be lifted when all $d$ orbitals have hybridization. Interestingly, apart from the usual repulsion, the levels manifest recombination features associated with the double phase transitions, for example, at large $\epsilon_p=9$ in upper panel. 

Moreover, the inset shows that the ``energy gap'' defined as $\Delta E \equiv E_1-E_g$ has almost exactly the same evolution between two models at relatively weak $t_z$ while apparently deviate at large $t_z$ after the phase transition. We found that the major difference of GS between two models lies in this large $t_z$ regime. Precisely, all $d$ orbital interlayer hybridization induces the system to avoid the GS degeneracy except at the exact critical points, where the GS nature is found to interchange with the first excited state corresponding to $E_1$ before the transition. 
In contrast, the model with only $d_{z^2}$ hybridization tends to result in a degenerate GS after the transition although at $\epsilon_p=9$ the transition is not approached in the shown range of $t_z$ (see also Fig.~\ref{phase2}).

Furthermore, as discussed earlier, all $d$ orbital hybridization results in exactly the same weights between a particular state and its parity state, e.g. $a_1L$-$b_1L$ and $b_1L$-$a_1L$ despite that the GS has no degeneracy. Nonetheless, in the presence of only $d_{z^2}$-$d_{z^2}$ hybridization, this weight degeneracy is replaced by the degeneracy of GS itself, where for each degenerate GS a state has different weight from its parity state; while this weight difference is reversed in the other degenerate GS. Therefore, the inversion symmetry breaking has distinct forms in two models.
This remarkable difference hidden in the GS degeneracy reflects the major role played by those $d$ orbitals other than $d_{z^2}$, whose additional interlayer hybridizations have smaller magnitude compared with $d_{z^2}$ (see our model for details), which is essential to generate a realistic model while tends to break the GS degeneracy.

\subsection{Hetero-bilayer: all $d$-orbital hybridization}

\begin{figure} [t!]
\center
\psfig{figure=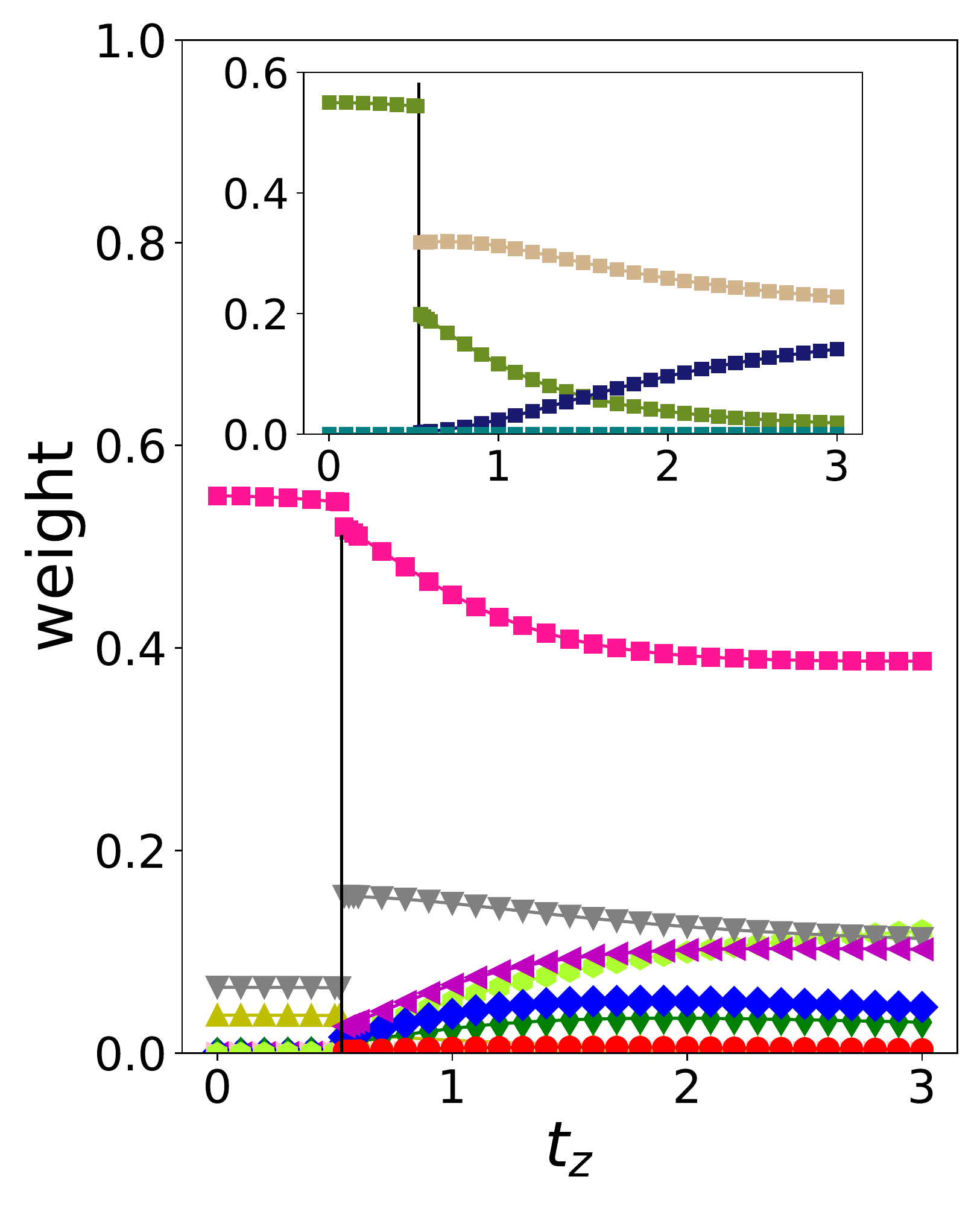},height=8.5cm,width=.48\textwidth, clip} 
\psfig{figure={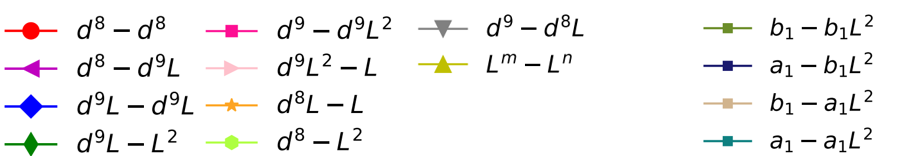},height=1.5cm,width=.48\textwidth, clip} 
\caption{GS weight evolution relevant for Cu$^{2+}$O$_2$-Ni$^{+}$O$_2$ with $\epsilon_p=3,7$ in two layers respectively. The dominant state corresponds to one hole on the layer with larger $\epsilon_p=7$ and three holes on the other layer with $\epsilon_p=3$ after charge transfer between layers.}	
\label{ep37}
\end{figure}

According to the above discussion on the homo-bilayer systems, the most interesting observation lies in the   local inversion symmetry breaking between layers. 
Now we switch to the hetero-bilayer systems, where the inversion symmetry has been artificially broken. These hetero-structure problem have long been a fruitful platform displaying versatile phenomena, ranging from interface superconductivity to 2D materials with artificial controllability.
In this work, we restrict our attention to the hetero-bilayer models with different $\epsilon_p$ and we show that the discrepancy between layers induces drastic impact on the nature of GS in terms of the charge transfer between layers. Fig.~\ref{ep37} provides an example of GS weight evolution relevant for Cu$^{2+}$O$_2$-Ni$^{+}$O$_2$ with $\epsilon_p=3,7$ for two layers respectively, although its experimental realization can be quite challenging given that the synthesis of infinite-layer nickelates has already encountered subtle difficulties~\cite{2019Nature,Botana_review}.

In the presence of asymmetric $\epsilon_p$ between layers, the holes preferentially locate on the layer with smaller $\epsilon_p$ such that there are states like one hole on one layer and three holes on the other, which is indeed confirmed as the dominant state $d^9$-$d^9L^2$ in Fig.~\ref{ep37}. The detailed composition is shown in the inset, which indicates that the single hole prefers $b_1$ orbital owing to its effective lowest site energy while the three-hole states transfers from $b_1L^2$ at relatively small $t_z$ into $a_1L^2$ state after the phase transition. Hence, once again we observe the gradual dominant role of $d_{z^2}$ orbital in the hetero-bilayer physics. Note that the critical $t_z=0.5$ is much smaller than that for homo-bilayer cases, which probably arises from the large heterogeneity between two layers.
Besides, the subdominant weights come from the three-hole $d^8L$, whose weight boosts after the transition originating from the same reasoning that $t_z$ hybridization promotes the triplet $d^8$ state similar to the homo-bilayers.

\begin{figure*}[t!]
\psfig{figure={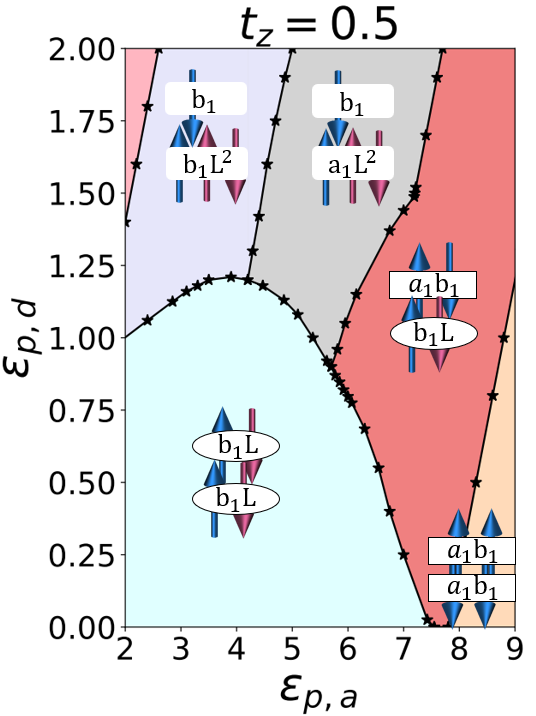},
height=6.5cm,width=.34\textwidth,angle=0,clip}
\psfig{figure={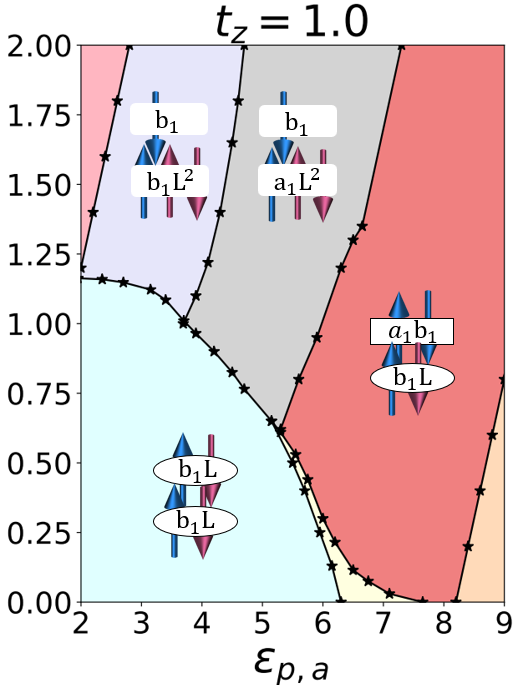},
height=6.5cm,width=.32\textwidth,angle=0,clip} 
\psfig{figure={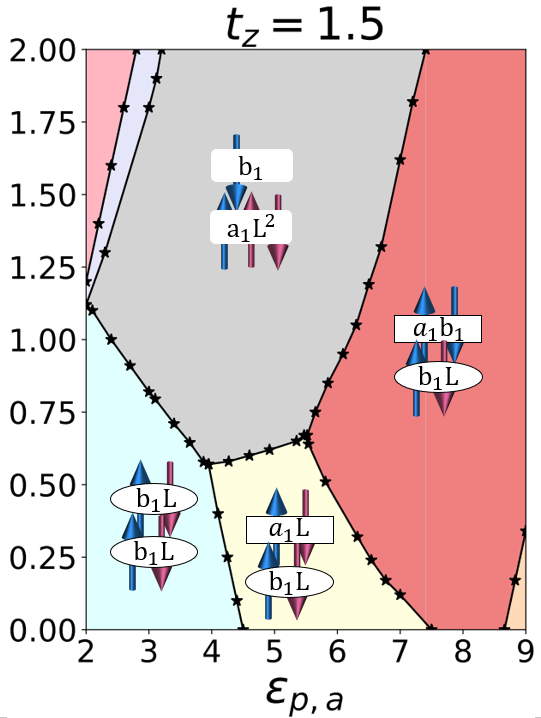},
height=6.5cm,width=.32\textwidth,angle=0,clip}
\psfig{figure={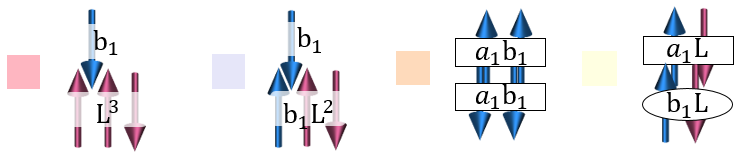},height=1.8cm,width=.6\textwidth, clip}
\caption{Phase diagram of hetero-bilayer of all $d$ hybridization with average $\epsilon_{p,a}=(\epsilon_{p1}+\epsilon_{p2})/2$ and difference $\epsilon_{p,d}=|\epsilon_{p1}-\epsilon_{p2}|/2$ for three typical $t_z=0.5,1.0,1.5$.}
\label{phasehetero}
\end{figure*}

Fig.~\ref{ep37} only illustrates an example with considerable asymmetry between $\epsilon_p$ in two layers. In order to systematically explore the impact of the $\epsilon_p$ discrepancy between layers and its interplay with interlayer hybridization, Fig.~\ref{phasehetero} displays the $\epsilon_{p,a}$-$\epsilon_{p,d}$ phase diagram with average $\epsilon_{p,a}=(\epsilon_{p1}+\epsilon_{p2})/2$ and difference $\epsilon_{p,d}=|\epsilon_{p1}-\epsilon_{p2}|/2$ for three typical $t_z=0.5,1.0,1.5$. For hetero-bilayer systems, we only studied the situation of all $d$ hybridization. 

Obviously, at weak hybridization $t_z=0.5$ (panel a), starting from the limiting homo-bilayer situation at $\epsilon_{p,d}=0$ (see also Fig.~\ref{phase1}), increasing the discrepancy $\epsilon_{p,d}$ generically locates the system into the states with one and three holes in each layer separately via the charge transfer from the layer with larger $\epsilon_p$ into the other layer with smaller $\epsilon_p$.  
However, this charge transfer transition disappears at sufficiently large $\epsilon_{p,a}$ reflecting the Hund's rule promoted $d^8$ triplet state within layers.

As $t_z$ increases, as shown in panels b and c, the regime of $b_1L$-$b_1L$ phase shrinks and partly replaced by $a_1L$-$b_1L$ phase (yellow regime) at small $\epsilon_{p,d}$, which is consistent with the above discussion in Fig.~\ref{phase1} on the homo-bilayer at $\epsilon_{p,d}=0$. Simultaneously, at large discrepancy $\epsilon_{p,d}$, the regime of $b_1$-$b_1L^2$ gives way to $b_1$-$a_1L^2$ as expected by the role of strong hybridization favoring $d_{z^2}$ orbital component reminiscent of homo-bilayers.

One particular interesting case relevant for Cu$^{2+}$O$_2$-Ni$^{+}$O$_2$ as shown in Fig.~\ref{ep37} corresponding to $\epsilon_{p,a}=5, \epsilon_{p,d}=2$, which lies in the critical region for small $t_z$ (panels a and b) while stronger interlayer hybridization transits its state readily into $b_1$-$a_1L^2$ regime. Our simplified four-hole simulation reveal that the charge transfer across the layers destabilizes the Zhang-Rice singlets within each layer. Together with the increasing significance of $a_1$ orbital, this can induce strong impact on the superconductivity in the realistic Cu$^{2+}$O$_2$-Ni$^{+}$O$_2$ interface or heterostructure.
In this context, more realistic many-body simulations are certainly desired to uncover more complete physical picture associated with this hetero-bilayer system.

\section{Conclusion}\label{Conclusion}

In summary, we used exact diagonalization to investigate a numerically tractable bilayer multi-orbital Hubbard model in the framework of our previous impurity approximation~\cite{Mi2020,Mi2020a} by incorporating the 3d$^{8}$ multiplet structure coupled to a full O-2p band. Aiming to describe the local electronic structure of hole doped homo-bilayer with the same $\epsilon_p$ and hetero-bilayer with distinct $\epsilon_p$, these simplified models display rich phase diagrams in both situations. To focus on the bound states relevant to the impurities and avoid the holes locating far from the two impurities in two layers, we restrict our attention to the case with interlayer hybridization between impurity's $d$ orbitals. 
Moreover, motivated by the recent experimental demonstration on the significance of the H doping~\cite{H nature}, we also investigated a even simpler model with only $d_{z^2}$-$d_{z^2}$ interlayer hybridization to account for the hybridization between the doped H and its nearest $d_{z^2}$ orbital so that there will be an effective interlayer $d_{z^2}$-$d_{z^2}$ hybridization albeit with much weaker magnitude than the situation without H blocking. 

For the homo-bilayer system, both models with all $d$-$d$ and $d_{z^2}$-$d_{z^2}$ hybridizations result in inversion symmetry breaking between two layers at strong enough interlayer hybridization, especially the coexistence of two-hole singlet and triplet states in opposite layers.  Besides, the interlayer hybridization generically tends to promote the significance of $d_{z^2}$ orbital. 
However, the examination of the GS degeneracy indicates the important difference between two models. Specifically, the $d$-$d$ hybridization leads to non-degenerate GS while the sole $d_{z^2}$-$d_{z^2}$ hybridization always induce degenerate GS. 

Given that the homo-bilayer model has potential connection to the high-pressure experiments on cuprate and even infinite-layer nickelates, our study uncovers the importance of multi-$d$ orbital description in the high-pressure setup. In addition, the mixture of singlet and triplet in separate layers potentially imply its detriment to the bulk superconductivity~\cite{Sunliling2022}, although more realistic consideration are certainly required.

For the hetero-bilayer system, inversion symmetry has been artificially broken. Our focus is the interplay between the discrepancy between two layers' $\epsilon_p$ and interlayer hybridization. 
In particular, we investigated a situation relevant for Cu$^{2+}$O$_2$-Ni$^{+}$O$_2$ with $\epsilon_p=3,7$ respectively, although its experimental realization can be quite challenging. 
Increasing the discrepancy $\epsilon_{p,d}$, namely the difference between $\epsilon_p$ of two layers, generally pushes the system into the states with one and three holes in each layer separately via the charge transfer from the layer with larger $\epsilon_p$ into the other layer with smaller $\epsilon_p$.  
However, this charge transfer transition disappears at sufficiently large $\epsilon_{p,a}$ reflecting the Hund's rule promoted $d^8$ triplet state within individual layers.
In addition, the gradual dominance of $d_{z^2}$ orbital also in the hetero-bilayer physics at sufficiently large interlayer hybridization is similar to the homo-bilayer systems. 

Our simplified four-hole simulation reveal the charge transfer across the layers so that destabilize the Zhang-Rice singlets within each layer. Together with the increasing significance of $a_1$ orbital, this can induce drastic impact on the superconductivity in the realistic Cu$^{2+}$O$_2$-Ni$^{+}$O$_2$ interface or heterostructure.
It would be quite interesting to explore whether the current observation of inversion symmetry breaking persists in the more physical but manageable many-body calculation on similar or the same type of bilayer multi-orbital models. Besides, the potential relevance to the bulk superconductivity in the context of bilayer or heterostructure between Cu$^{2+}$O$_2$ and Ni$^{+}$O$_2$, their relation to high-pressure settings, and also the significant role of $d_{z^2}$ orbital are worthwhile for further exploration~\cite{siliang,Bansil,Maier2019}.

\section{Acknowledgements} 
We acknowledge the useful discussion with Wenxin Ding, Zhenzhong Shi, Jian Kang, and Xiancong Lu. This work was supported by National Natural Science Foundation of China (NSFC) Grant No. 12174278, startup fund from Soochow University, and Priority Academic Program Development (PAPD) of Jiangsu Higher Education Institutions. 


\end{document}